\documentclass[aps,letterpaper,prc,twocolumn,showpacs,floatfix,nofootinbib,preprintnumbers,superscriptaddress,amsmath,amssymblinenumbers]{revtex4-2}
\usepackage{multirow}
\usepackage{stackengine}
\usepackage{subfigure}
\usepackage{hyperref}
\hypersetup{
    colorlinks=true,  %
    linkcolor=blue,
    urlcolor=blue,
    citecolor=blue
}
\usepackage{dsfont}
\usepackage{mathtools}
\let\plainRho\rho
\renewcommand{\rho}[1][0pt]{%
  \mathrel{\raisebox{3pt}{$\plainRho$}}%
}
\usepackage[table]{xcolor}
\usepackage{xurl}
\hypersetup{breaklinks=true}

\begin{document}
  \title{Halo structure of $^6$He from \textit{ab initio} two-nucleon spatial correlations}
  \author{Mengyao Huang}
  \thanks{Contact author: mengyaoh01@gmail.com}
  \thanks{This research was performed while Mengyao Huang was in graduate school at Iowa State University}
 
  \affiliation{Department of Physics and Astronomy, Iowa State University, Ames, Iowa 50011, USA}
  \affiliation{Lawrence Livermore National Laboratory, P.O. Box 808, L-414, Livermore, California 94551, USA}
  \author{Tobias Frederico}
  \affiliation{Instituto Tecnol{\'o}gico de Aeron{\'a}utica, DCTA, 12228-900 S{\~a}o Jos{\'e} dos Campos, Brazil}
  \author{Peng Yin}
  \affiliation{College of Physics and Engineering, Henan University of Science and Technology, Luoyang 471023, China}
  \affiliation{CAS Key Laboratory of High Precision Nuclear Spectroscopy, Institute of Modern Physics, Chinese Academy of Sciences, Lanzhou 730000, China}
  \affiliation{Department of Physics and Astronomy, Iowa State University, Ames, Iowa 50011, USA}
  \author{Robert~A.~M. Basili}
  \affiliation{Department of Physics and Astronomy, Iowa State University, Ames, Iowa 50011, USA}
  \affiliation{Department of Mathematics, Iowa State University, Ames, Iowa 50011, USA}
  \author{Patrick J. Fasano}
  \affiliation{Department of Physics and Astronomy, University of Notre Dame, Notre Dame, Indiana 46556, USA}
  \author{James P. Vary}
  \affiliation{Department of Physics and Astronomy, Iowa State University, Ames, Iowa 50011, USA}
  \begin{abstract}

We evaluate pairwise correlations using ground state wave functions for $^4$He and $^6$He obtained by \textit{ab initio} no-core shell model calculations with the Daejeon16 nucleon-nucleon interaction plus Coulomb interaction, to characterize the structures of these two systems. We demonstrate that two-nucleon spatial correlations, specifically the pair-number operator $r^0$ and the square-separation operator $r^2$ projected on two-body spin $S$ and isospin $z$ components encode important details of the halo structure of $^6$He. We also analyze the single-particle state occupancies and the two-body state occupancies for the ground state of $^4$He and $^6$He. Our results indicate that the two valence neutrons in the ground state of $^6$He dominantly form a spin-singlet configuration. The rms pair separations between core nucleons and halo neutrons of $^6$He are, on average, about 80\% larger than pair separations within the swollen and off-centered ``$\alpha$ core''. We show that this off-centering effect is primarily responsible for the observed increase in point-proton radius $r_p$ in $^6$He relative to $^4$He.

  \end{abstract}

  \maketitle

\section{Introduction}

Clustering phenomena in atomic nuclei reflect both the degrees of freedom of individual nucleons and collective structures within the nucleus. As a natural example, certain nuclei near the ``drip lines'' (where nucleons are nearly free to escape) are likely formed with a relatively stable core and a ``halo'' made of one or more nucleons loosely bound to the core \cite{zhukov_bound_1993,HansenPG1995NH,RevModPhys.76.215,doi:10.1063/1.1301722,Riisager_2013}.

Neutron halos have drawn significant attention since the 1980s when the surprisingly large interaction cross section of $^{11}$Li compared with other Li isotopes was observed in radioactive ion beam (RIB) transmission experiments \cite{PhysRevLett.55.2676,Tanihata_1996,tanihata_nuclear_2016}. These nuclei also display a narrow core momentum distribution in peripheral breakup reactions \cite{PhysRevLett.60.2599,PhysRevLett.69.2050}. Strong electric dipole ($E1$) excitations at low excitation energies were observed in some of the halo nuclei, attributed to their weakly bound halo neutron(s) undergoing excitation to the continuum states \cite{PhysRevLett.96.252502,PhysRevC.59.1252,PhysRevC.70.054606}. Although there is no clear-cut distinction between halo and nonhalo nuclei, halo nuclei are recognized by their shared characteristics. Nuclei with a neutron halo, for example, display a long tail in the neutron density distribution profile, and have small neutron separation energies for the halo neutron(s). 

Over the years, efforts have been made to extract the clustering information from \textit{ab initio} calculations. For instance, form factors calculated by quantum Monte Carlo \cite{PhysRevC.84.024319} were used to assess the clustering patterns. The $^4\mbox{He (ground state)}+n+n$ cluster dynamics were studied within the \textit{ab initio} no-core shell model (NCSM) with continuum \cite{PhysRevLett.117.222501,PhysRevC.97.034332}. Recently, measures of entanglement of $^4$He and $^6$He nuclei were explored using \textit{ab initio} nuclear many-body calculations \cite{PhysRevC.103.034325}, where a core-valence structure naturally emerges from the full no-core calculation of $^6$He by analyzing common features of entanglement.

From a computational point of view, studying the clustering properties of nuclei could help us design improved basis functions and usefully subdivide the Hilbert space into regions of greater and lesser importance. Spatial information such as the average relative separations of nucleons in a system is key to characterizing the cluster wave functions. Accordingly, we examine the structure of a nucleus by \textit{ab initio} calculations utilizing two-body correlations. Previously, one-body probabilities have been used in the \textit{ab initio} NCSM \cite{BARRETT2013131} with great success in revealing information on the clustering of Li isotopes \cite{PhysRevC.86.034325,cockrell2012ab}. Using two-body operators has the advantage of unfolding the detailed structure of the dominant configuration and displaying the differences in the nucleon-nucleon correlations among pairs of nucleons in specific two-body configurations. Besides the most commonly used spatial parameters (the rms radii), we explore the two-body spatial moments $r^0$ and $r^2$---called the pair-number operator and the square-separation operator, respectively---projected on each two-body state sector.

We choose $^6$He as a test case in this new method owing to its relative simplicity. It is the lightest Borromean halo nucleus, which means that neither of the two-component substructures, $n$-$n$ or $n$-$\alpha$ in the $\alpha$+$n$+$n$ system, is bound when one removes the third component. In this case, the one-body density profile cannot fully reveal the geometrical structure of the 2$n$-halo, while two-body correlations can probe the relative distance between the two valence neutrons compared with the relative distances of the other pairs of nucleons. Our primary focus is to introduce our method and demonstrate its application to $^6$He in advance of achieving fully converged \textit{ab initio} NCSM results for the two-body correlations that we develop and employ. We defer the development of suitable extrapolation methods for these two-body correlations to a later project.

In the following sections, we first introduce the coordinate space two-body operators we use (Sec.~\ref{operators}) and the approach to solving the quantum many-body problem (Sec.~\ref{NCSM}). We then present and discuss the calculated expectation values of these two-body operators and how they relate to the halo structure of $^6$He (Sec.~\ref{results}). 
\section{Constructing Two-body Operators in Coupled-\texorpdfstring{$J$}{J} Basis}\label{operators}

To carry out a quantitative analysis of the distributions of nucleons and pairs of nucleons within a nucleus, one needs appropriate operators that sample these quantities.  For this purpose, we define two-body operators whose expectation values within the nuclear ground state will inform our detailed analysis of how nucleons are arranged in coordinate space on average. %
The standard basis we use is the spherical single-particle basis, which is denoted by its quantum numbers, including radial quantum number $n$, orbital angular momentum $l$, spin $s$, total angular momentum $j$, total angular momentum projection $m_j$, isospin $t$ and isospin $z$-projection $t_{z}$. We can use the short-hand notation
\begin{equation}
\alpha = \{n_\alpha, l_\alpha, \frac{1}{2}, j_\alpha, m_{j_\alpha}, \frac{1}{2}, t_{z\alpha}\},
\end{equation}
where the first ``1/2'' is for nucleon spin $s$ and the second ``1/2'' is for nucleon isospin $t$.

To compute matrix elements of two-body operators in a basis with ``good'' two-body total angular momentum $J$, we work in terms of the spherical tensor coupled product \cite{Edmonds+2016}, e.g.,
\begin{equation}
\{a_\alpha^\dagger a_\beta^\dagger\}_{J,M} = \smashoperator{\sum_{m_{j_\alpha},m_{j_\beta}}}C\!
\begin{smallmatrix}
j_\alpha & \!\!j_\beta & \!\!J \\
m_{j_\alpha} & \!\!m_{j_\beta} & \!\!M
\end{smallmatrix}
\,a_\alpha^\dagger a_\beta^\dagger,
\end{equation}
where $M$ is the $z$-projection of $J$. $C\!\begin{smallmatrix}
j_\alpha & \!\!j_\beta & \!\!J \\
m_{j_\alpha} & \!\!m_{j_\beta} & \!\!M
\end{smallmatrix}$ are the Clebsch-Gordan coefficients. 
To keep notation compact, we suppress the label $M$ 
and introduce the resulting coupled-$J$ basis states, 
\begin{equation}
|\alpha\beta;J\rangle=\{a_\alpha^\dagger a_\beta^\dagger\}_{J}|0\rangle.
\end{equation}

When calculating observables represented by one-body and two-body operators, one may compute the corresponding density matrices between the desired initial and final states, from which all matrix elements can subsequently be extracted via weighted summation. The two-body density matrix in the coupled-$J$ basis is
\begin{widetext}
\begin{equation}\label{rho}
\begin{split}
\rho_{\{\alpha\beta\}_{J'}\{\gamma\delta\}_{J}}&=\langle\Phi'_{A}|\{a_\alpha^\dagger a_\beta^\dagger\}_{J'} \{a_\delta a_\gamma\}_{J}|\Phi_{A}\rangle\\
&=\smashoperator{\sum_{m_{j_\alpha},m_{j_\beta}}} C\!\begin{smallmatrix}
j_\alpha &\!\! j_\beta &\!\! J' \\
m_{j_\alpha} &\!\! m_{j_\beta} &\!\! M'
\end{smallmatrix}
\smashoperator{\sum_{m_{j_\gamma},m_{j_\delta}}} C\!
\begin{smallmatrix}
j_\gamma &\!\! j_\delta &\!\! J \\
m_{j_\gamma} &\!\! m_{j_\delta} &\!\! M
\end{smallmatrix}
\rho_{\alpha\beta\gamma\delta}
\end{split}
\end{equation}
where $\rho_{\alpha\beta\gamma\delta}$ are two-body density matrix elements in the uncoupled single-particle representation.

The expectation value of a two-body operator in a system of nucleons expressed in the coupled-$J$ basis can thus be written as
\begin{equation}
\langle \Phi'_{A}|\mathcal{O}_\text{2B}|\Phi_A\rangle = \smashoperator{\sum_{\{\alpha\beta\}_{J'}\{\gamma\delta\}_{J}}} \rho_{\{\alpha\beta\}_{J'}\{\gamma\delta\}_{J}}\langle \alpha\beta;J' |\mathcal{O}|\gamma\delta;J\rangle,
\end{equation}
where ordered indices ($\alpha<\beta,\gamma<\delta$) are implicitly assumed wherever summing over pairs of two-body states and $\langle \alpha\beta;J' |\mathcal{O}|\gamma\delta;J\rangle$ is antisymmetrized under particle exchange.

A rotationally invariant two-body operator $\mathcal{O}(r)$ suitable for our purpose, with a given two-body spin $S$ and a given pair of isospin $z$ projections $\{t_{za},t_{zb}\}$, acting on the coupled-$J$ basis, yields the matrix elements (derived in Appendix~\ref{rotscal})

\begin{equation}
\begin{split}
\langle \alpha\beta;J'M'|\mathcal{O}(r)[S,t_{za},t_{zb}]|\gamma\delta;J M\rangle =&\,\delta_{t_{z\alpha},t_{z\gamma}}\delta_{t_{z\beta},t_{z\delta}}\delta_{J'J}\delta_{M'M}\,g(t_{z\alpha},t_{z\beta})_{t_{za}\leq t_{zb}}\\
&\times(2S+1)\sqrt{(2j_\alpha + 1)(2j_\beta + 1)(2j_\gamma + 1)(2j_\delta + 1)}\\
&\times\sum_{L }(2L+1)
\begin{Bmatrix}
l_\alpha & l_\beta & L\\
1/2 & 1/2 & S\\
j_\alpha & j_\beta & J
\end{Bmatrix}
\begin{Bmatrix}
l_\gamma & l_\delta & L\\
1/2 & 1/2 & S\\
j_\gamma & j_\delta & J
\end{Bmatrix}\langle n_\alpha l_\alpha n_\beta l_\beta L ||O(r)||n_\gamma l_\gamma n_\delta l_\delta L \rangle,
\end{split}
\end{equation}
where $L$ is the two-body total orbital angular momentum and we define $g(t_{z\alpha},t_{z\beta})_{t_{za}\leq t_{zb}}\equiv \delta_{t_{z\alpha},t_{za}}\delta_{t_{z\beta},t_{zb}}+(1-\delta_{t_{za},t_{zb}})\delta_{t_{z\alpha},t_{zb}}\delta_{t_{z\beta},t_{za}}$ which has the property:
\begin{equation}\label{match_isospin}
g(t_{z\alpha},t_{z\beta})_{t_{za}\leq t_{zb}} = \begin{cases}
1, &\text{if $t_{z\alpha}=t_{za}$ and $t_{z\beta}=t_{zb}$}\\
1, &\text{if $t_{z\alpha}=t_{zb}$ and $t_{z\beta}=t_{za}$}\\
1, &\text{if $t_{z\alpha}=t_{z\beta}=t_{za}=t_{zb}$}\\
0, &\text{otherwise}
\end{cases}.
\end{equation}
\end{widetext}
This function ensures that any pair of isospin $z$ projections $\{t_{z\alpha},t_{z\beta}\}$ has at most one way to match the given values $\{t_{za},t_{zb}\}$ with $t_{za}\leq t_{zb}$, as well as the exchange symmetry of $t_{z\alpha}$ and $t_{z\beta}$.

We then introduce a notation to describe an operator acting within a system of nucleons (similar to the $U[u]$ and $V[v]$ operators defined in Ref.~\cite{Caprio_2020})---the subscript $A$, which means that we apply the operator on all nucleon pairs in the $A$-body space, i.e. $\mathcal{O}_A
\equiv\sum_{i<j}^A\mathcal{O}$ for two-body operators and $\mathcal{O}_A
\equiv\sum_{i}^A\mathcal{O}$ for one-body operators. Thus, the expectation value of $\mathcal{O}(r)[S,t_{za},t_{zb}]$ in the $A$-body space is
\begin{equation}\label{br}
\begin{split}
\langle\mathcal{O}(r)_{A}[S,t_{za},t_{zb}]\rangle
=&\left\langle\sum_{i<j}^A\mathcal{O}(r_{ij})[S,t_{za},t_{zb}]\right\rangle\\
=&\smashoperator{\sum_{\{\alpha\beta\}_{J'}\{\gamma\delta\}_{J}}} \rho_{\{\alpha\beta\}_{J'}\{\gamma\delta\}_{J}}\\
&\times\langle \alpha\beta;J'|\mathcal{O}(r)[S,t_{za},t_{zb}]|\gamma\delta;J\rangle,
\end{split}
\end{equation}
where $r_{ij}=|\Vec{r}_i-\Vec{r}_j|$ is the relative distance between the $i$th and $j$th nucleon, and $i,j=1,2,\ldots, A$. $\alpha$ and $\gamma$ ($\beta$ and $\delta$) denote the initial and final single-particle states of the $i$th ($j$th) nucleon, respectively. In the following, the subscript $A$ is conveniently omitted when referring to the expectation values of the operator (denoted by a bracket directly around the operator), i.e., $\langle\mathcal{O}\rangle\equiv\langle\mathcal{O}_A\rangle$. 

In Sec.~\ref{pi_o} and Sec.~\ref{ss_o}, we define the pair-number operator and the square-separation operator with given $S$ and isospin $z$ projections, which determine the nucleon species (proton-proton, neutron-proton and neutron-neutron, indicated as $pp$, $pn$, and $nn$, respectively). Combining them will provide information on mean square separations of a $pp$, $pn$ or $nn$ pair of nucleons with a given two-body spin $S$, which will be illustrated in Sec.~\ref{exp_ss_o}.

\subsection{The pair-number operator}\label{pi_o}

The pair-number operator $r^0$ for a given two-body spin $S$ and a fixed pair of values of the isospin $z$-projections is $r^0[S,t_{za},t_{zb}]\equiv P_SP_{t_{za},t_{zb}}$. When applied to the $A$-nucleon space, its expectation value $\langle r^0[S,t_{za},t_{zb}] \rangle$ is
\begin{equation}\label{r0}
\smashoperator{\sum_{\{\alpha\beta\}_{J}\{\gamma\delta\}_{J}}} \rho_{\{\alpha\beta\}_{J}\{\gamma\delta\}_{J}}\!\langle \alpha\beta;J|P_SP_{t_{za},t_{zb}}|\gamma\delta;J\rangle.
\end{equation}

$\langle r^0[t_{za},t_{zb}] \rangle$ is obtained by summing over $S$ in the expression above,
\begin{equation}\label{r0_tatb_sum}
\begin{split}
&\smashoperator{\sum_{\{\alpha\beta\}_{J}\{\gamma\delta\}_{J}}} \rho_{\{\alpha\beta\}_{J}\{\gamma\delta\}_{J}}\!\!\langle \alpha\beta;J|P_{t_{za},t_{zb}}|\gamma\delta;J\rangle\\
=&\smashoperator{\sum_{\{\alpha\beta\}_{J}}}\rho_{\{\alpha\beta\}_{J}\{\alpha\beta\}_{J}}g(t_{z\alpha},t_{z\beta})_{t_{za}\leq t_{zb}},
\end{split}
\end{equation}
which is an integer according to Eq.~(\ref{match_isospin}) and (\ref{rho}), meaning the number of pairs with a given pair of isospin $z$ projections (however, $\langle r^0[S,t_{za},t_{zb}] \rangle$ is generally not an integer). It will become clearer when we present the calculation results in Sec.~\ref{chapter_r0}.

Taking into account all possible isospin $z$ projections,
\begin{equation}\label{r0_sum}
\langle r^0 \rangle=\smashoperator{\sum_{t_{za}\leq t_{zb}}}\langle r^0[t_{za},t_{zb}] \rangle
=\sum_{i<j}^A 1=\frac{A(A-1)}{2},
\end{equation} 
which is the total number of pairs of nucleons in an $A$-nucleon system. The value of 1 inside the summation of the second line of Eq.~(\ref{r0_sum}) comes from that any pair of nucleons' isospin $z$-projection values automatically falls into one of the first three cases in Eq.~(\ref{match_isospin}) when there is no restriction on the values of isospin $z$ projections.

\subsection{The square-separation operator}\label{ss_o}

The expectation value of the square-separation operator $r^2$ in the two-body space describes the relative distance squared for a pair of nucleons. The square-separation operator for a given two-body spin $S$ and a fixed pair of values of isospin $z$ projections is defined as $r^2[S,t_{za},t_{zb}]\equiv r^2P_SP_{t_{za},t_{zb}}$. Its expectation value in an $A$-body space $\langle r^2[S,t_{za},t_{zb}] \rangle$ is
\begin{equation}\label{r2ep}
\smashoperator{\sum_{\{\alpha\beta\}_{J}\{\gamma\delta\}_{J}}} \rho_{\{\alpha\beta\}_{J}\{\gamma\delta\}_{J}}\!\langle \alpha\beta;J|r^2P_S P_{t_{za},t_{zb}}|\gamma\delta;J\rangle.
\end{equation}

$\langle r^2[t_{za},t_{zb}]\rangle$ represents the expectation value for restrictions on isospin $z$ projections only, which is
\begin{equation}
\smashoperator{\sum_{\{\alpha\beta\}_{J}\{\gamma\delta\}_{J}}} \rho_{\{\alpha\beta\}_{J}\{\gamma\delta\}_{J}}\!\langle \alpha\beta;J|r^2|\gamma\delta;J\rangle \,g(t_{z\alpha},t_{z\beta})_{t_{za}\leq t_{zb}}.
\end{equation}

\subsection{Relation to the rms radii}
The square of the nuclear matter rms radius $r_m$ with center-of-mass coordinates $\Vec{R}_{\text{cm}}=(\sum_i^A\Vec{r}_i)/A$ is
\begin{equation}\label{rm}
r^2_{m} \equiv\frac{1}{A}\sum_i^A(\Vec{r}_i-\Vec{R}_{\text{cm}})^2=\frac{1}{A^2}\sum_{i<j}^A(\Vec{r}_i-\Vec{r}_j)^2=\frac{r_A^2}{A^2},
\end{equation}
where $r_A^2$ is the square-separation operator in the $A$-nucleon space. Projecting $r^2_A$ on different isospin $z$ components and scaled by $1/A^2$ yields the point-proton and point-neutron rms radii $r_p$ and $r_n$ \cite{PhysRevC.90.034305}. This means that the convergence behavior of the results from the $r^2$ operator is similar to that of rms radii. Thus, in Sec.~\ref{sec_rms_radii}, we use calculated rms radii to determine the suitable parameters for our calculation of $r^0$ and $r^2$.

\section{\textit{Ab initio} NCSM Calculation}\label{NCSM}

We obtain the matrix elements of operators $r^0$ and $r^2$ in the harmonic oscillator (HO) relative motion basis (Appendix~\ref{appendix_ho_matrix}), and 
transform them to two-body matrix elements in the single-particle basis by the Moshinsky transformation sketched in Appendix~\ref{appendix_moshinsky}. The resulting two-body matrix elements are then employed to calculate expectation values using many-body NCSM wavefunctions. We adopt the Daejeon16 two-nucleon ($N\!N$) interaction \cite{SHIROKOV201687} as well as Coulomb interaction in our \textit{ab initio} NCSM calculations using the MFDn code \cite{MARIS201097,mfdn2,SHAO20181,fasano_2025_18013362}. The Daejeon16 interaction is obtained by softening the chiral effective field theory-based ($\chi$EFT-based) Idaho next-to-next-to-next-to-leading order (N3LO) $N\!N$ interaction using the similarity renormalization group (SRG) method \cite{Wegner1994,GlazekWilson1994}. Five significant figures are preserved throughout our calculation process.

The calculated rms radii for $^4$He and $^6$He are presented in Sec.~\ref{sec_rms_radii}. They 
provide a handle for choosing the many-nucleon basis space cutoff $N_\text{max}$ and HO energy spacing $\hbar\Omega$ to minimize the basis truncation error of the expectation values for the $r^0$ and $r^2$ operators with limited computational resources. We then compare these rms radii with the experimental values to 
underpin the utility
of the calculations and inform the uncertainties 
arising from our NCSM calculations.
Later, we present in Sec.~\ref{chapter_r0} the results of the $r^0$ operator and analyze the results in detail with the help of the occupation distribution of single-particle states and two-body states, which also result from 
our calculations. We show how the $r^0$ and $r^2$ operators provide insight into the halo structure of $^6$He in Sec.~\ref{chapter_r0} and Sec.~\ref{exp_ss_o}.

\section{Results and Analysis}\label{results}

\subsection{The rms radii}\label{sec_rms_radii}
\begin{figure*}[htp!]
\centering
\includegraphics[width=\linewidth]{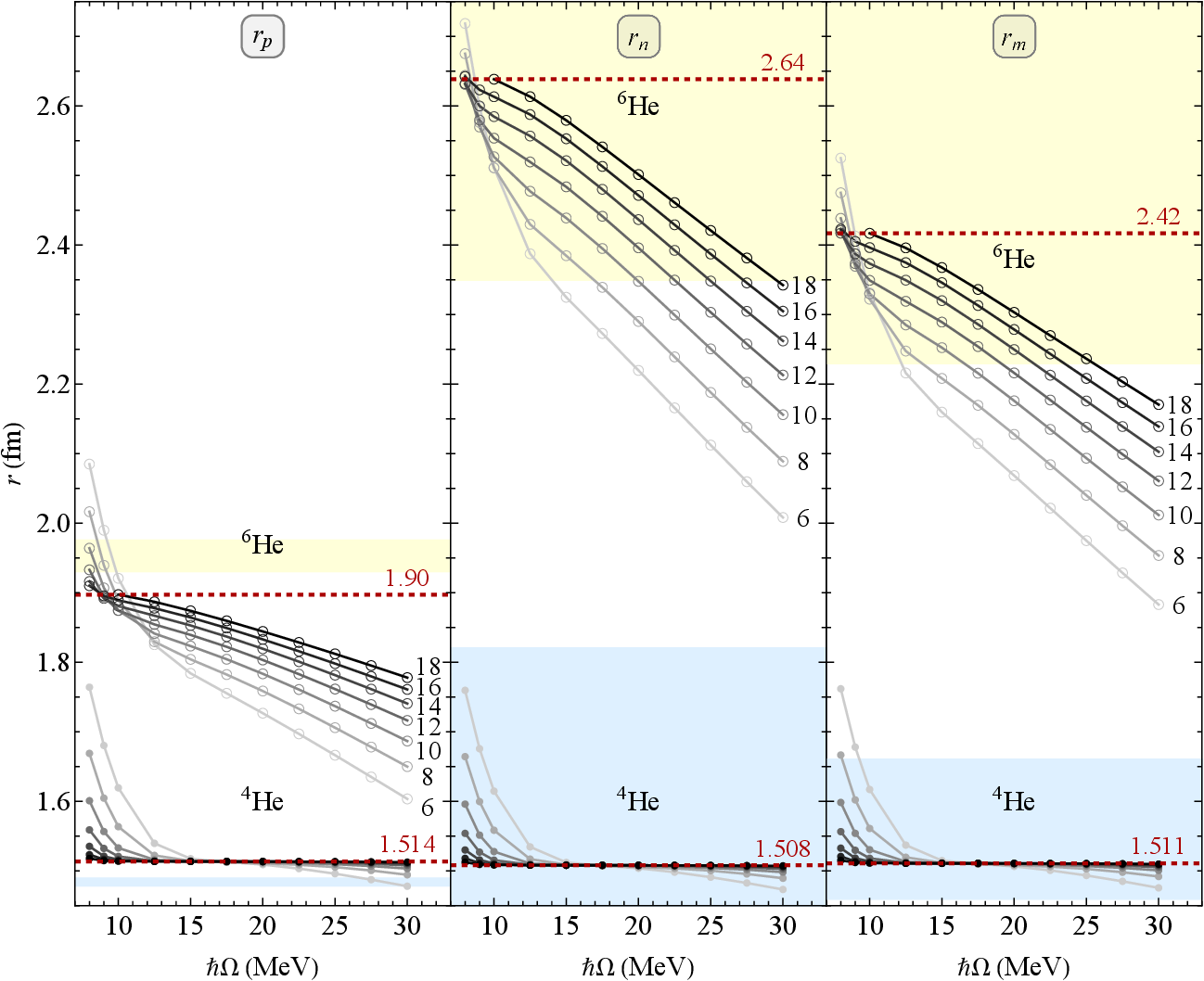}
\caption{$r_p$, $r_n$ and $r_m$ of $^4$He ($^6$He) with the Daejeon16 interaction plus Coulomb potential, calculated using \textit{ab initio} NCSM. Results for $^4$He ($^6$He) are represented by filled (open) circles, which are connected by straight lines for the same $N_\text{max}$, with darker lines for higher $N_\text{max}$ ($N_\text{max}$ values are labeled on the right of the $^6$He results). The horizontal red dashed lines indicate $r^*_p$, $r^*_n$ and $r^*_m$ as defined in the text. The experimental ranges are indicated by blue (for $^4$He) and yellow (for $^6$He) bands.}
\label{fig:rms_radiusHe4He6}
\end{figure*}

To determine $N_{\text{max}}$ and $\hbar\Omega$ values that
limit our truncation error, 
we calculate $r_p$, $r_n$ and $r_m$ on a grid of $N_{\text{max}}$ from 6 to 18 with a step size of 2, and $\hbar\Omega$ from 10 to 30~MeV, sampled every 2.5 
MeV with additional points at 8 and 9 MeV for $N_{\text{max}} \leq 16$.
The results for $^4$He and $^6$He are shown in Fig.~\ref{fig:rms_radiusHe4He6}. 

\begin{figure}[htp!]
\centering
\includegraphics[width=\linewidth]{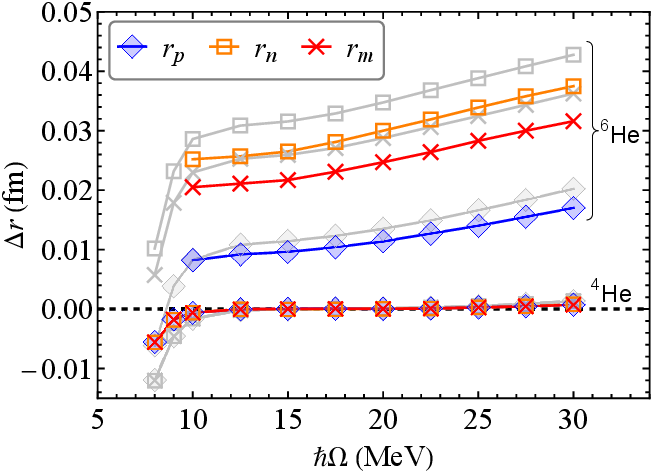}
\caption{The difference in $r_p$, $r_n$, $r_m$ between consecutive $N_\text{max}$ values: $N_\text{max}=18$ values minus $N_\text{max}=16$ values (color) and $N_\text{max}=16$ values minus $N_\text{max}=14$ values (gray). The absolute value of the difference $|\Delta r|$ serves as an indicator of the basis truncation uncertainty of $r_p$, $r_n$, $r_m$ at $N_\text{max}=16$.}
\label{fig:uncertainty_lower_bound_He4He6}
\end{figure}

We find the range $\hbar\Omega=10$ to 27.5~MeV provides an approximate $N_{\text{max}}$ independence for the point-nucleon rms radii for $^4$He.
This is evident 
in Fig.~\ref{fig:uncertainty_lower_bound_He4He6} through the very small differences (on the order of 10$^{-4}$ fm) in the point-nucleon rms radii between $N_{\text{max}}=16$ and $N_{\text{max}}=18$ over this range of $\hbar\Omega$. For $^6$He, however, the differences between neighboring $N_{\text{max}}$ (denoted as $\Delta r$) decrease with increasing $N_{\text{max}}$ but do not display a very good convergence pattern, 
as shown in both Fig.~\ref{fig:rms_radiusHe4He6} and \ref{fig:uncertainty_lower_bound_He4He6}. 
With an emphasis on minimizing $\Delta r$ for $^6$He while achieving a reasonable 
$\hbar\Omega$-independence for it, we adopt $\hbar\Omega=10$~MeV. This choice also provides good convergence of $\Delta r$ for $^4$He.
Finally, we adopt $N_{\text{max}}=16$ since it is sufficient for our demonstration purposes of calculating our suite of characteristic matrix elements with limited computational resources. 

We show in Table~\ref{rms_r} $r_p$, $r_n$, and $r_m$ for both $^4$He and $^6$He with $N_{\text{max}}=16$ and $\hbar\Omega=10$ MeV. We note that $r_n$ is about 39\% (0.7~fm) larger than $r_p$ for $^6$He. On the other hand, $r_p$ increases by $25\%$ from $^4$He (1.515~fm) to $^6$He (1.889~fm). As we will show, this increase in $r_p$ may be understood as 
arising from the motion of the charged ``$\alpha$ core'' around the common center of mass, as well as from contributions from swelling of the $\alpha$ core \cite{RevModPhys.85.1383}. Specifically, we show in Sec.~\ref{exp_ss_o} that the rms separations indicate that the swelling effect only contributes a few percent increase to $r_p$ while the center-of-mass deviation effect is prominent in $^6$He.

\begin{table}[t] \centering
\renewcommand{\arraystretch}{1.2}
\setlength{\extrarowheight}{1pt}
\setlength{\tabcolsep}{0.5pt}
\caption{$r_p$, $r_n$ and $r_m$ [fm] for both $^4$He and $^6$He with $N_\text{max}=16$ and $\hbar \Omega=10$ MeV. $r^*_p$, $r^*_n$ and $r^*_m$ [fm] are for $N_\text{max}=18$ with $\hbar \Omega$ corresponding to the minimum gap between $N_\text{max}=18$ and $N_\text{max}=16$ results, serving as a reference for the uncertainty owing to the $N_\text{max}$ truncation. The experimental values $r_p^\text{expt}$, $r_n^\text{expt}$ and $r_m^\text{expt}$ [fm] are reported, where $r_p^\text{expt}$ and $r_m^\text{expt}$ are separately measurable, while $r_n^\text{expt}$ is obtained using the relation $Ar^2_m=Zr^2_p+Nr^2_n$.}
\label{rms_r}
\begin{tabular}{@{\extracolsep{2.5pt}}ccccccccc@{}}
\hline \hline
& \multicolumn{3}{c}{$^4$He} & \multicolumn{3}{c}{$^6$He}
\\
\cline{2-4}\cline{5-7}
$x$ 
& $r_x$
& $r^*_x$
& $r_x^\text{expt}$
& $r_x$
& $r^*_x$
& $r_x^\text{expt}$
 \\ \hline 
$p$ & 1.5145 & 1.5136 & 1.484(5) \cite{PhysRevC.77.041302} & 1.8891 & 1.8973 & 1.953(22) \cite{PhysRevLett.108.052504}\\
$n$ & 1.5091 & 1.5081 & 1.43--1.82 & 2.6132 & 2.6384 & 2.35--3.08 \\
$m$ & 1.5118 & 1.5109 & 1.46--1.66 \cite{RevModPhys.85.1383} & 2.3963 & 2.4168 & 2.23--2.75 \cite{RevModPhys.85.1383}
 \\ \hline \hline
\end{tabular}
\end{table} 

To analyze the uncertainties resulting from the $N_\text{max}=16$ truncation and examine the consistency with the experimental values, we present $r^*$ and $r^\text{expt}$ values in Table~\ref{rms_r}, explained as follows: $r_p^*$, $r_n^*$ and $r_m^*$ are those obtained with $N_{\text{max}}=18$ and $\hbar\Omega$ corresponding to the minimum gap $\Delta r$ between $N_{\text{max}}=18$ and $N_{\text{max}}=16$ results (For $^4$He, $\hbar\Omega=15$~MeV for $r^*_p$ and $r^*_m$, and $\hbar\Omega=17.5$~MeV for $r^*_n$; For $^6$He, $\hbar\Omega=10$~MeV). Comparing $r$ and $r^*$ values gives an estimate of the magnitude of the basis cutoff uncertainties: For $^4$He, the differences appear in the third decimal digit; For $^6$He, the differences show up in the second decimal digit. The experimental point-nucleon rms radii $r_p^\text{expt}$, $r_n^\text{expt}$, and $r_m^\text{expt}$ are indicated in the last column in Table~\ref{rms_r}. $r_p^\text{expt}$ is calculated using $r_p^2=r_c^2-R_p^2-3/(4M_p^2)-(N/Z)R_n^2-r_{so}^2$, where the nucleus rms charge radius $r_c$ can be measured from laser spectroscopy or scattering experiments, the Darwin-Foldy term $3/(4M_p^2)=0.033$~fm$^2$ \cite{PDG2020} and the neutron mean square charge radius $R_n^2=-0.1161(22)$ \cite{PDG2020}. The relativistic correction due to spin-orbit interaction $r_{so}^2=0$ for $^4$He and $-$0.08~fm$^2$ for $^6$He \cite{PhysRevC.82.014320}. 
The latest recommended value of $R_p=0.84075(64)$ \cite{RevModPhys.97.025002}. %
$r_m^\text{expt}$ in Table~\ref{rms_r} is obtained from the interaction cross sections of elastic scattering \cite{RevModPhys.85.1383}, while $r_n^\text{expt}$ is computed from the universal relation $Ar^2_m=Zr^2_p+Nr^2_n$. The experimental ranges in Table~\ref{rms_r} are visualized in Fig.~\ref{fig:rms_radiusHe4He6} as color bands. Due to the large experimental uncertainties, $r^*_m$ ($r^*_n$) remains consistent with $r_m^\text{expt}$ ($r_n^\text{expt}$) even without full convergence. On the other hand, the fully converged $r^*_p$ for $^4$He exceeds the upper edge of the experimental values by 0.02~fm, which is mainly due to the properties of the Daejeon16 potential \cite{PhysRevC.106.034305}. The nonconverged $r^*_p$ for $^6$He is 0.03~fm smaller than the lower edge of the experimental values, and 
it has a favorable trend
with larger $N_\text{max}$. In summary, comparing $r$ with $r^*$, we estimate that the truncation error is 0.001~fm (0.03~fm) for $^4$He ($^6$He). Comparing $r^*$ with $r^\text{expt}$, we estimate that the systematic error (due to the potential) is 0.02~fm.

We continue with $N_\text{max}=16$ and $\hbar \Omega=10$~MeV to calculate the expectation values for the $r^0$ and $r^2$ operators. We present six (five) decimal digits for the results of $\langle r^0[S,t_{za},t_{zb}] \rangle$ ($\langle r^2[S,t_{za},t_{zb}] \rangle$) to prevent losing significant digits in later calculations.

\subsection{The expectation values of the $r^0$ operator}\label{chapter_r0}

$\langle r^0[S,t_{za},t_{zb}] \rangle$ represents the (likely fractional) number of two-nucleon systems that carry a given $S$ and a specific pair of isospin projection values. The results of $\langle r^0[S,t_{za},t_{zb}] \rangle$ for both values of $S$ and for all the $\{t_{za},t_{zb}\}$ combinations are presented in Table~\ref{table_r0}.

\begin{table}[htp] 
\centering
\renewcommand{\arraystretch}{1.2}
\setlength{\extrarowheight}{1pt}
\setlength{\tabcolsep}{8pt}
\caption{$\langle r^0[S,t_{za},t_{zb}]\rangle$ calculated to five significant digits (denoted ``calc.''), with extra digits shown to prevent round-off error in later calculations. The naive expected value (denoted ``naive'') of each calculated result, based on occupying only the lowest available states in an HO basis, is listed for comparison as discussed in the text.}
\label{table_r0}
\begin{tabular}{@{\extracolsep{4pt}}cccccc@{}} 
\hline\hline
 &  & \multicolumn{2}{c}{$^4$He} & \multicolumn{2}{c}{\addstackgap[2pt]{$^6$He}}  \\ 
\cline{3-4}\cline{5-6}
 & S & naive & calc. & naive & calc.\\
\hline 
$pp$ & 0 & 1 & 0.976772 & 1 & 0.967578 \\
$pp$ & 1 & 0 & 0.023228 & 0 & 0.032422 \\
$pn$ & 0 & 1 & 0.979111 & 2 & 1.977430 \\
$pn$ & 1 & 3 & 3.020890 & 6 & 6.022570 \\
$nn$ & 0 & 1 & 0.976774 & 3 & 2.865420 \\
$nn$ & 1 & 0 & 0.023226 & 3 & 3.134580 \\ 
\hline 
\multicolumn{2}{c}{all} & 6 & 6.000001 & 15 & 15.00000 \\ 
\hline\hline
\end{tabular}
\end{table}

First, we perform a sanity check for the sum of $\langle r^0[S,t_{za},t_{zb}] \rangle$ over all $\{S,t_{za},t_{zb}\}$ combinations for $^4$He and $^6$He (see Table~\ref{table_r0}) to confirm that they are equal to the total number of nucleon pairs $A(A-1)/2$ for each nucleus as indicated by Eq.~(\ref{r0_sum}). We note that although the calculated $\langle r^0[S,t_{za},t_{zb}]\rangle$ are not integers, a sum of $\langle r^0[S,t_{za},t_{zb}] \rangle$ over two-body spin $S$ equals the number of pairs with given $\{t_{za},t_{zb}\}$ in $^4$He and $^6$He, as suggested in Eq.~(\ref{r0_tatb_sum}). For example, $\langle r^0[0,1/2,1/2]\rangle_{^4\text{He}}+\langle r^0[1,1/2,1/2]\rangle_{^4\text{He}}=0.976772+0.023228=1.000000=\langle r^0[pp]\rangle_{^4\text{He}}$, recovering an integer within 5 significant digits.

$^4$He results are very close to integers and to naive expectations. We expect and find that both $pp$ and $nn$ in the ground state of $^4$He are dominantly spin-singlet because this is the only $l=0$ basis state ($s$ state) respecting the overall antisymmetrization for exchanging of two isospin-symmetric fermions, i.e. $(-1)^{l+S+T}=-1$, where $l$, $S$, and $T$ are the two-body orbital angular momentum in the two-body relative frame (see Appendix~\ref{appendix_moshinsky}), two-body spin, and two-body isospin, respectively. One may also expect, as we indeed find, a small but non-zero portion of spin-triplet states (0.023228 for $pp$ $S=1$ and 0.023226 for $nn$ $S=1$) arising from partial wave contributions with higher values of $l$. The added Coulomb potential makes the isospin symmetry only approximate in our calculations (note that we assume equal mass for $p$ and $n$, which is taken as the average mass of these nucleons; the Daejeon16 interaction is charge-independent), which contributes to the small difference between $pp$ and $nn$ expectation values for $^4$He. %

\begin{figure*}[t]
\centering
  \begin{tabular}[c]{c @{\hspace{1ex}} c}
    \includegraphics[height=4.9cm]{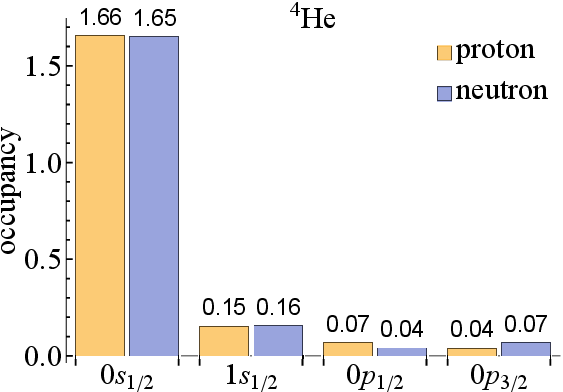}
    &
    \includegraphics[height=4.9cm]{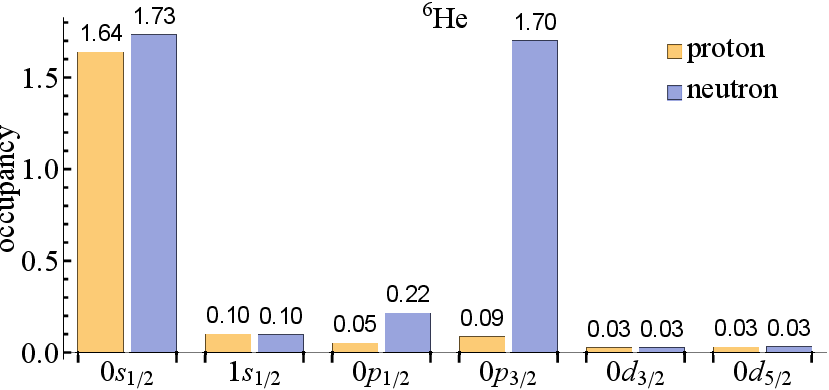}
    \\
    (a) & (b)
  \end{tabular}
\caption{Occupancy of protons and neutrons in HO single-particle states for the ground state of $^4$He (a) and $^6$He (b), with $N_\text{max}=16$ and $\hbar\Omega=10$ MeV. The HO single-particle state occupancies that are major contributors are shown and aggregate to 
more than $90\%$ of the proton number or neutron number of each nucleus.}
\label{fig:he4he6_probability}
\end{figure*}

\begin{table}[htbp]
\caption{(a) Single-particle occupancies obtained by occupying nucleons on lowest Pauli-allowed HO states. (b) Number of nucleon pairs in each two-body charge-dependent category. For $^6$He, we further divide neutrons into $\alpha$ core ($n^\alpha$) and halo ($n^h$) contributions.}
\label{tab:two-body_categories}
\begin{ruledtabular}
\begin{tabular}{l c c c}
 & $^4$He & $^6$He \\
\hline
(a) & & & \\
$p$ & 2 & 2 & \\
\multirow{2}{*}{$n$} & \multirow{2}{*}{2} & $n^\alpha$ & 2\\
& & $n^h$ & 2\\
all & 4 & 6 & \\
(b) & & & \\
$pp$ & 1 & 1 & \\
\multirow{2}{*}{$pn$} & \multirow{2}{*}{4} & $pn^\alpha$ & 4 \\
& & $pn^h$ & 4 \\
\multirow{3}{*}{$nn$} & \multirow{3}{*}{1} & $n^\alpha n^\alpha$ & 1 \\
& & $n^\alpha n^h$ & 4 \\
& & $n^h n^h$ & 1 \\
all & 6 & 15 & \\
\end{tabular}
\end{ruledtabular}
\end{table}

For $pn$ of $^4$He in Table~\ref{table_r0}, the result for spin-triplet is approximately triple the result for spin-singlet for the reason explained below. For the spin-independent parts of the $N\!N$ interactions, singlet and triplet two-body spin states have the same energy in the same HO wave function of relative motion (labeled by $n$ and $l$ for two-body central potentials)---there is no preference among the allowed two-body spin configurations $\uparrow\uparrow$ (triplet), $\downarrow\downarrow$ (triplet), $\uparrow\downarrow+\downarrow\uparrow$ (triplet), and $\uparrow\downarrow-\downarrow\uparrow$ (singlet). As a result, the ratio of the probability of triplet to singlet two-body spin states for the ground state of $^4$He $pn$ is approximately 3:1. Upon considering the spin-dependent part of the $N\!N$ potential, e.g., $S_pS_n$ coupling in the $s$ state, which can be usefully written as $V_\text{spin}=\hbar^{-2} V_1(r)\Vec{S}_p\cdot\Vec{S}_n$ \cite{ns_mit:2012}, the probability of finding a pair of $pn$ in one of the triplet states will be slightly higher than in the singlet state if $V_1(r)$ is effectively attractive (thus negative), as in deuterium. This assessment takes into account $\langle \Vec{S}_p\cdot\Vec{S}_n\rangle=+\hbar^2/4$ for triplet states and $-3\hbar^2/4$ for singlet states. These conditions are met, as can be seen in the calculated $pn$ results of $^4$He in Table~\ref{table_r0}.

To pave the way for detailed inspection of the 
$\langle r^0[S,t_{za},t_{zb}]\rangle$ results conveyed in Table~\ref{table_r0}, we consider the NCSM results for the occupancies of the lowest HO single-particle states in $^4$He and $^6$He as shown in Fig.~\ref{fig:he4he6_probability}.  The dominant occupancies are seen to be near the limits of a naive shell-model picture where the lowest Pauli-allowed states are occupied assuming the traditional spin-orbit splitting of p-shell states.  
In particular, Fig.~\ref{fig:he4he6_probability} shows that the lowest HO $s$ state 
contains well above 80\% of the maximum occupation number of two for protons and neutrons in both $^4$He and $^6$He. In addition, for $^6$He, we obtain a similar occupation for neutrons in the HO $0p_{3/2}$ orbit. With more than 80\% of nucleons occupying the lowest Pauli-allowed HO states, we adopt the naive shell-model picture as a starting point for investigating matrix elements of two-nucleon operators. 

In Table~\ref{tab:two-body_categories}, we present the extreme values for HO single particle state occupancies [part (a)] and the expectation values of pair-number operator $r^0$ [part(b)] as a function of charge for $^4$He and $^6$He. For neutrons in $^6$He, we further divide their contributions into the $s$-state components which we call the $\alpha$ core neutrons (denoted $n^{\alpha}$) and the $p$-state components or halo neutrons (denoted $n^h$).
We regard $n^\alpha$ and $n^h$ as two categories of neutrons; they then form two categories for $pn$ ($pn^\alpha$ and $pn^h$) and three categories for $nn$ ($n^\alpha n^\alpha$, $n^\alpha n^h$ and $n^h n^h$), as shown in Table~\ref{tab:two-body_categories}(b).

\begin{figure*}[t]
\centering
  \begin{tabular}[c]{c @{\hspace{1ex}} c @{\hspace{0ex}} c}
    \includegraphics[height=4.22cm]{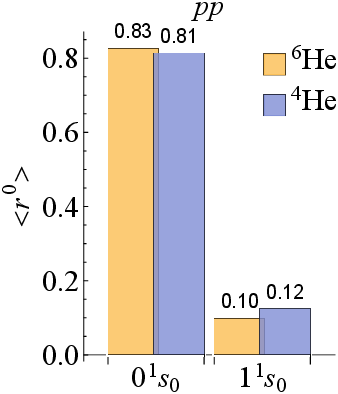}
    &
    \includegraphics[height=4.22cm]{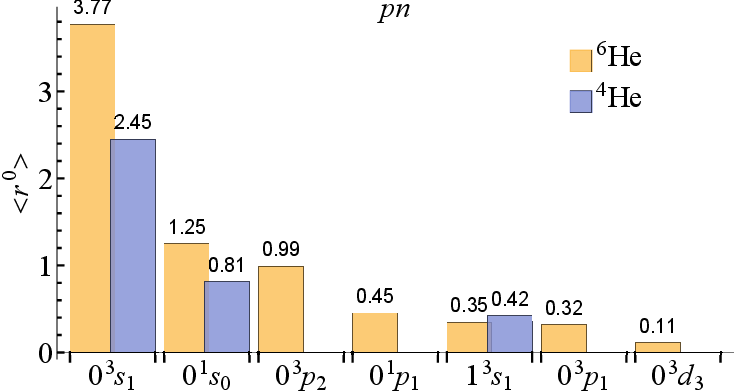}
    &
    \includegraphics[height=4.22cm]{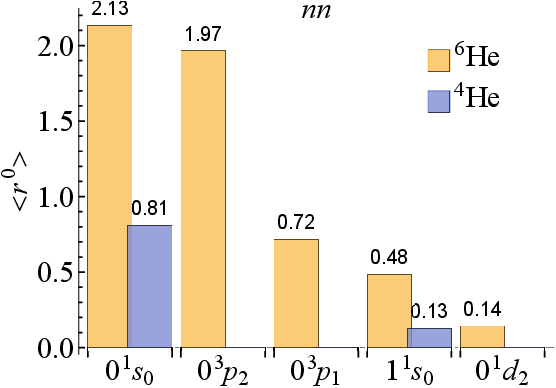}
    \\
    (a) & (b) & (c)
  \end{tabular}
\caption{Major $\langle r^0\rangle$ two-body contributions of the ground state of $^4$He and $^6$He with $N_\text{max}=16$ and $\hbar\Omega=10$ MeV, for (a) $pp$, (b) $pn$, and (c) $nn$ pairs according to the HO state of relative motion. The bins are arranged by decreasing magnitude for $^6$He, with the corresponding $^4$He bins shown alongside for comparison. The states shown above account for more than $90\%$ of $\langle r^0\rangle$ in each isospin-projected pair configuration.}
\label{fig:he4he6_r0}
\end{figure*}

The analysis above helps to see why the calculated $\langle r^0 [S,t_{za},t_{zb}]\rangle$ of $^6$He in Table~\ref{table_r0} are also consistent with naive expectations. 
In particular, since there is only a small change of proton occupancy from $^4$He to $^6$He, we expect and find that $pp$-pair occupancy should remain largely in the spin-singlet state. Correspondingly, 
$pn^\alpha$ has similar correlations as $pn$ of $^4$He. In other words, it is thrice as likely to find a spin-triplet state than to find a spin-singlet state in $pn^\alpha$. 

We speculate that the ratio of spin-triplet to spin-singlet state occupancies in $pn^h$ will be close to those of $pn^\alpha$. The reason is as follows: First, for the $^6$He neutrons, the $0s_{1/2}$ and $0p_{3/2}$ occupancies are large and approximately equal. Second, the condition that the total angular momentum is zero suggests that the $n^\alpha$ and $n^h$ have the same spin distributions.
Therefore, $pn^h$ will likely mimic $pn^\alpha$ with nearly no preference among the four two-body spin states: $\uparrow\uparrow$ (triplet), $\downarrow\downarrow$ (triplet), $\uparrow\downarrow+\downarrow\uparrow$ (triplet), and $\uparrow\downarrow-\downarrow\uparrow$ (singlet). For the same reason, except for $p$ replaced by $n^\alpha$ in $pn^h$, the two-body spin states for $n^\alpha n^h$ will mimic those for $pn^{\alpha}$, i.e., three $n^\alpha n^h$ pairs in spin-triplet states and one $n^\alpha n^h$ pair in a spin-singlet state. To a lesser extent, we also expect both $n^\alpha n^\alpha$ and $n^h n^h$ to be dominantly in spin-singlet states, assuming that the correlations of neutrons within each shell are stronger than the correlations of neutrons across different shells. This assumption seems to be approximately true, because the difference between the expected and calculated $\langle r^0 [S,t_{za},t_{zb}]\rangle$ for $^6$He $nn$ pairs in Table~\ref{table_r0} is only 0.13. 

We now address the role of $N_\text{max}$ truncation. The maximum difference in $\langle r^0 [S,t_{za},t_{zb}]\rangle$ between $N_\text{max}=16$ and $N_\text{max}=14$ is $10^{-4}$ for $^6$He $pp$ and $pn$ pairs and $5\times 10^{-4}$ for $^6$He $nn$ pairs. 
These are meaningful differences since 
the MFDn calculation preserves five significant digits.
On the other hand, for $^4$He, the maximum difference in $\langle r^0 [S,t_{za},t_{zb}]\rangle$ between $N_\text{max}=16$ and $N_\text{max}=14$ is $4\times 10^{-5}$, so the uncertainty due to truncation that we quote only appears in the last significant decimal digit.

Let us now examine the two-body state contributions to $\langle r^0 \rangle$ (labeled by spectroscopic notation $n\,^{2S+1}l_J$, where $|l-S|\leq J\leq |l+S|$) in Fig.~\ref{fig:he4he6_r0} for the three isospin-projected pair configurations. For the $^4$He ground state, all the major components come from $s$ states, as expected from a naive shell model. $^4$He $pp$ is mostly ($81\%$) in the lowest spin-singlet state, with a small portion ($12\%$) in the $n=1$ spin-singlet $s$ state. $^6$He $pp$ is almost identical to $^4$He. $^4$He $pn$ spin-triplet states are dominated by the lowest two $s$ states ($96\%$), while the $^4$He $pn$ spin-singlet state is dominated by the lowest $s$ state ($81\%$). For $^6$He $pn$ pairs, we see the rise of $p$ and $d$ states in both spin-singlet and spin-triplet configurations. The three major $p$ states contribute $22\%$ and the $0\prescript{3}{}{d}_3$ state contributes $1\%$ to the overall $\langle r^0\rangle$. For $nn$, we again see the rise of higher states ($45\%$ $p$ states and $2\%$ $0\prescript{1}{}{d}_2$) in $^6$He. We note that, for $nn$, due to the $(-1)^{l+S+T}=-1$ antisymmetry constraint, all allowed $p$ states are spin-triplets and all allowed $s$ and $d$ states are spin-singlets. We observe that $^6$He $nn$ spin triplets are dominantly in $p$ states and $^6$He $nn$ spin singlets are dominantly in $s$ states.

To sum up, we conclude that the two $^6$He valence neutrons dominantly form a singlet spin configuration and are mostly in $s$ states, from the consideration of preserving the $^4$He two-body spin configuration inside $^6$He [suggested by the fact that the proton and neutron occupation probabilities at the $0s_{1/2}$ single-particle state in Fig.~\ref{fig:he4he6_probability} and $pp$ two-body spin states in Fig.~\ref{fig:he4he6_r0}(a) remain almost unchanged from $^4$He to $^6$He], and from the fairly good consistency between the \textit{ab initio} NCSM-calculated $\langle r^0\rangle$ two-body spin and isospin projections and the naively expected values (Table~\ref{table_r0}). This provides an indicator of halo structure in $^6$He through the following reasoning: The $nn$ spin-singlet $s$ states have symmetric spatial wave functions $\Psi_{nlm}$ which have nonzero probability close to the center-of-mass of the pair due to the absence of a centrifugal term in the relative Schr\"{o}dinger equation, resulting in stronger correlation between the two nucleons compared with spin-triplet states. This seems to support a ``dineutron'' halo at some distance from the center-of-mass of the nucleus. The ``cigar''-like configuration is another possibility seen in many calculations but appears not to be the dominant configuration \cite{zhukov_bound_1993,PhysRevC.72.044321,PhysRevC.84.051304}.

It can be seen from the above analyses that the \textit{ab initio} NCSM-calculated $\langle r^0 \rangle$ two-body spin and isospin projections are useful for speculating about the structure of the  $^6$He ground state. We show in Sec.~\ref{exp_ss_o} that combining $\langle r^0 \rangle$ with $\langle r^2 \rangle$ gives us a more quantitative picture of the structure. %

\subsection{The expectation values of the $r^2$ operator}\label{exp_ss_o}

The expectation values of the $r^2$ operator for different two-body spin and isospin $z$-projections $\langle r^2[S,t_{za},t_{zb}] \rangle$ are presented in Table~\ref{table_r2}. To normalize $\langle r^2[S,t_{za},t_{zb}] \rangle$ to its average value for a single pair of nucleons, we define the mean square separation between two nucleons of given $\{S,t_{za},t_{zb}\}$ as
\begin{equation}\label{r2_S_tz}
\overline{\langle r^2[S,t_{za},t_{zb}] \rangle} \equiv
 \frac{\langle r^2[S,t_{za},t_{zb}]\rangle}{\langle r^0[S,t_{za},t_{zb}]\rangle},
\end{equation}
and the rms separation between two nucleons of given $\{S,t_{za},t_{zb}\}$ as
\begin{equation}\label{r_S_tz}
\overline{\langle r^2[S,t_{za},t_{zb}] \rangle}^{1/2} \equiv
 \sqrt{\frac{\langle r^2[S,t_{za},t_{zb}]\rangle}{\langle r^0[S,t_{za},t_{zb}]\rangle}}.
\end{equation}
These derived quantities are shown in the last two columns for each nucleus in Table~\ref{table_r2}. 

\begin{table}[htp] \centering
\renewcommand{\arraystretch}{1.2}
\setlength{\extrarowheight}{1pt}
\setlength{\tabcolsep}{1.4pt}
\caption{$\langle r^2[S,t_{za},t_{zb}]\rangle$ [fm$^2$] with five significant digits, with extra digits shown to prevent round-off error in later calculations. We also present the mean square separation $\overline{\langle r^2[S,t_{za},t_{zb}]\rangle}$ [fm$^2$] and the rms separation $\overline{\langle r^2[S,t_{za},t_{zb}]\rangle}^{1/2}$ [fm] as defined in the text.}
\label{table_r2}
\begin{tabular}{@{\extracolsep{4pt}}cccccccc@{}}
\hline\hline
&  & \multicolumn{3}{c}{$^4$He} & \multicolumn{3}{c}{\addstackgap[2pt]{$^6$He}} \\
\cline{3-5}\cline{6-8}
 & S
 & $\langle r^2\rangle$
 & $\overline{\langle r^2\rangle}$
 & $\overline{\langle r^2\rangle}^{1/2}$
 & $\langle r^2\rangle$
 & $\overline{\langle r^2\rangle}$
 & \addstackgap[2pt]{$\overline{\langle r^2\rangle}^{1/2}$}\\
\hline
$pp$ & 0 & 5.95862 & 6.10032 & 2.46988 & 6.63201 & 6.85424 & 2.61806 \\ 
$pp$ & 1 & 0.22754 & 9.79581 & 3.12983 & 0.36145 & 11.1484 & 3.33892 \\
$pn$ & 0 & 5.94563 & 6.07248 & 2.46424 & 24.5046 & 12.3921 & 3.52024 \\ 
$pn$ & 1 & 18.3172 & 6.06352 & 2.46242 & 73.2384 & 12.1607 & 3.48721 \\
$nn$ & 0 & 5.89396 & 6.03411 & 2.45644 & 38.7177 & 13.5121 & 3.67588 \\
$nn$ & 1 & 0.22648 & 9.75110 & 3.12268 & 63.2643 & 20.1827 & 4.49252 
 \\ \hline \hline
\end{tabular}
\end{table}

These results can be interpreted with the following considerations. Using again the overall antisymmetrization property of fermion pairs, as mentioned in Sec.~\ref{chapter_r0}, the rms separation for spin-singlet is expected to be smaller than that for spin-triplet for the $pp$ and $nn$ pairs in their lowest partial waves. This turns out to be true for both $^4$He (2.5~fm versus 3.1~fm) and $^6$He (2.6~fm versus 3.3~fm for $pp$; 3.7~fm versus 4.5~fm for $nn$) in Table~\ref{table_r2}. For $pn$ pairs, since the two-body isospin $T$ can be either 0 or 1, the spatial part of the wave function for a given $S$ can be either symmetric or antisymmetric. Thus, the rms separations for spin-singlet $pn$ pairs and spin-triplet $pn$ pairs are similar (2.46 fm versus 2.46 fm for $^4$He; 3.52 fm versus 3.49 fm for $^6$He).

Comparing $^6$He with $^4$He, we see that the rms separation for the dominant $pp$ spin-singlet configuration is expanded by 6\% from $^4$He to $^6$He. Interestingly, the remaining ${\approx}3\%$ (Table~\ref{table_r0}) portion of $pp$ spin-triplet states also experiences a similar expansion in rms separation from $^4$He to $^6$He.
On the other hand, the rms separations for $nn$ and $pn$ pairs are expanded by $40\%{\sim}50\%$ from $^4$He to $^6$He, consistent with the increase of $r_n$ and $r_m$ from $^4$He to $^6$He shown in Fig.~\ref{fig:rms_radiusHe4He6} and Table~\ref{rms_r}.

Summing over $S$ in both the denominator and numerator of Eq.~(\ref{r2_S_tz}), we define the mean square separation for two nucleons of given $\{t_{za},t_{zb}\}$ as
\begin{equation}
\overline{\langle r^2 [t_{za},t_{zb}]\rangle} \equiv
 \frac{\sum\limits_{S}\langle r^2[S,t_{za},t_{zb}]\rangle}{\sum\limits_{S}\langle r^0[S,t_{za},t_{zb}]\rangle}=\frac{\langle r^2[t_{za},t_{zb}]\rangle}{\langle r^0[t_{za},t_{zb}]\rangle}.
\end{equation}
$^4$He and $^6$He have only one $pp$ pair, therefore $\langle r^0[1/2,1/2]\rangle_{^4\text{He},^6\text{He}}=1$. Similarly, it is straightforward to note that $\langle r^0[-1/2,1/2]\rangle_{^4\text{He}}=4$, $\langle r^0[-1/2,1/2]\rangle_{^6\text{He}}=8$, $\langle r^0[-1/2,-1/2]\rangle_{^4\text{He}}=1$, and $\langle r^0[-1/2,-1/2]\rangle_{^6\text{He}}=6$. The rms separation of given $\{t_{za},t_{zb}\}$ can be defined as the square root of $\overline{\langle r^2 [t_{za},t_{zb}]\rangle}$. The calculated mean square separations along with the rms separations are shown in Table~\ref{table_r2_sumS}. We compare them with the corresponding results computed using NNLO$_\text{opt}$ nucleon-nucleon interaction \cite{PhysRevLett.110.192502} with $N_\text{max}=14$ in Ref.~\cite{saaf2016bridging} and find overall consistency.

\begin{table*}[htp] 
\centering
\renewcommand{\arraystretch}{1.2}
\setlength{\extrarowheight}{1pt}
\setlength{\tabcolsep}{2pt}
\caption{$\langle r^2[t_{za},t_{zb}] \rangle$ [fm$^2$] computed from Table~\ref{table_r2}, together with the mean square separation $\overline{\langle r^2[t_{za},t_{zb}]\rangle}$ [fm$^2$] and the nucleon-nucleon rms separation $\overline{\langle r^2[t_{za},t_{zb}]\rangle}^{1/2}$ [fm]. We compare our rms separations with those of Ref. \cite{saaf2016bridging}. }
\label{table_r2_sumS}
\begin{tabular}{@{\extracolsep{4pt}}ccccccccc@{}}
\hline \hline
& \multicolumn{4}{c}{$^4$He} & \multicolumn{4}{c}{\addstackgap[2pt]{$^6$He}}
\\
\cline{2-5}\cline{6-9}
 & $\langle r^2\rangle$
 & $\overline{\langle r^2\rangle}$
 & $\overline{\langle r^2\rangle}^{1/2}$
 & \cite{saaf2016bridging}
 & $\langle r^2\rangle$
 & $\overline{\langle r^2\rangle}$
 & \addstackgap[1.5pt]{$\overline{\langle r^2\rangle}^{1/2}$}
 & \cite{saaf2016bridging}
 \\ \hline 
$pp$ & 6.1862 & 6.1862 & 2.4872 & 
2.40 &
6.9935 & 6.9935 & 2.6445 &
2.62 \\
$pn$ & 24.263 & 6.0657 & 2.4629 & 
2.40 &
97.743 & 12.218 & 3.4954 &
3.3 \\ 
$nn$ & 6.1204 & 6.1204 & 2.4740 & 
2.40 &
101.98 & 16.997 & 4.1227 &
3.9
 \\ \hline \hline
\end{tabular}
\end{table*}

To characterize a nucleus by an overall nucleon-nucleon separation without specifying nucleon species, we define the mean square separation between two nucleons in a nucleus as
\begin{equation}\label{r2_A}
\overline{\langle r^2 \rangle} \equiv
 \frac{\sum\limits_{S,t_{za}\leq t_{zb}}\langle r^2[S,t_{za},t_{zb}]\rangle}{\sum\limits_{S,t_{za}\leq t_{zb}}\langle r^0[S,t_{za},t_{zb}]\rangle}
 =\frac{\langle r^2\rangle}{A(A-1)/2},
\end{equation}
where $A(A-1)/2=6$ for $^4$He and 15 for $^6$He. Then the nucleon-nucleon rms separation is the square root of $\overline{\langle r^2 \rangle}$. From Table~\ref{table_r2} and Eq.~(\ref{r2_A}), $\overline{\langle r^2 \rangle}_{^4\text{He}}=6.0949$~fm$^2$ and $\overline{\langle r^2 \rangle}_{^6\text{He}}=13.781$~fm$^2$, which yield $\overline{\langle r^2 \rangle}^{1/2}_{^4\text{He}}=2.4688$~fm and $\overline{\langle r^2 \rangle}^{1/2}_{^6\text{He}}=3.7123$~fm, meaning that the nucleon-nucleon separations in $^6$He are, on average, 50\% larger than those in $^4$He.

In summary, the classical structure of $^4$He is a tetrahedron, analyzed by combining rms radii with rms separations. From Table~\ref{rms_r}, the calculated point-proton and point-neutron rms radii for $^4$He are nearly identical, around $r_0=1.51$~fm. From Table~\ref{table_r2_sumS}, $^4$He $pp$, $pn$, and $nn$ rms separations are also very similar, about $a=2.47$~fm. The ratio ${a}/{r_0}=1.636$ is approximately the same as the side length to the center-to-vertex distance ratio of a tetrahedron ${2\sqrt{6}}/{3}= 1.633$. This suggests a route by which we can relate our quantum many-body results to a classical picture that depicts $^4$He as a tetrahedron with protons and neutrons at the tetrahedron's vertices.

To analyze the structure of $^6$He, we notice a 6.3\% increase in $pp$ rms separation from $^4$He to $^6$He (Table~\ref{table_r2_sumS}), representing a substantially weaker effect than the corresponding 25\% increase in $r_p$ (Table~\ref{rms_r}). This means that not only does the $\alpha$ core swell, but also the center of mass of $^6$He is no longer centered at the center of the $\alpha$ core. The center of mass is likely to have longer distances from the two protons in $^6$He, compared with $^4$He. Meanwhile, $pn$ ($nn$) rms separation of $^6$He increases by 41.9\% (66.6\%), compared with that of $^4$He. Since $pn$ and $nn$ contain more than one pair of nucleons, a binary model is constructed to capture the main features of these changes. We introduce $x$ ($y$) as the ratio between the rms separation of $pn^\alpha$ or $n^{\alpha}n^{\alpha}$ ($pn^h$, $n^{\alpha}n^{h}$, or $n^{h}n^{h}$) in $^6$He and the rms separation of $pn$ or $nn$ in $^4$He, i.e.,
\begin{equation}
  \frac{r_{pn \text{ or } nn}^{(^6\text{He})}}{r_{pn \text{ or } nn}^{(^4\text{He})}}=\left\{
    \begin{array}{ll}
      x, & \mbox{both $^6$He nucleons within the $\alpha$ core}\\
      y, & \mbox{otherwise}
    \end{array}
  \right..
\end{equation}
Matching the percentage increases,
\begin{equation}
\frac{r_{pn}^{(^6\text{He})}-r_{pn}^{(^4\text{He})}}{r_{pn}^{(^4\text{He})}}=\sqrt{\frac{4x^2+4y^2}{8}}-1=41.9\%,
\end{equation}
and
\begin{equation}
\frac{r_{nn}^{(^6\text{He})}-r_{nn}^{(^4\text{He})}}{r_{nn}^{(^4\text{He})}}=\sqrt{\frac{x^2+5y^2}{6}}-1=66.6\%.
\end{equation}
Solving these two equations yields $x=0.93$ and $y=1.78$. This suggests that the $pn$ or $nn$ rms separation within the $\alpha$ core stays nearly the same as for $^4$He ($x\approx 1$), while the rms separation of two nucleons with correlation extending outside of the $\alpha$ core is found to increase by almost 80\%, supporting the formation of a halo.

\section{Conclusion}\label{conclusion}

We show that the formation of the halo structure of $^6$He can be examined from the coordinate-space two-body correlation operators $r^0$ and $r^2$ using the \textit{ab initio} NCSM with the Daejeon16 plus Coulomb interactions.

We examine the dominant two-body spin configurations of $^4$He and $^6$He by analyzing the results of $r^0$ projected on two-body spin and isospin $z$ components. Combined with the analysis of the single-particle states and the two-body states of the ground state of $^4$He and $^6$He, the two valence neutrons of $^6$He predominantly form a spin-singlet configuration where symmetry regarding the probability of forming an arbitrary two-body spin is largely respected. We demonstrate that the inclusion of results for expectation values of $r^2$ provides a more subtle picture of $^4$He and $^6$He via the geometrical relations among the calculated point-nucleon rms radii and the nucleon-nucleon rms separations. We utilize a binary model to describe the nucleon-nucleon separations of $^6$He. We conclude that $^4$He classical structure is essentially a tetrahedron, and the nucleon-nucleon rms separation within the $\alpha$ core of $^6$He is comparable to that of $^4$He. On the other hand, the rms separation between the core nucleons and the halo neutrons of $^6$He is about 80\% larger than the above scale. We observe the signature of a moderately swollen $^6$He $\alpha$ core through the 6.3\% increase in rms separation between the two protons compared with $^4$He. Since this cannot explain the 25\% increase in the point-proton radius $r_p$ of $^6$He relative to $^4$He, we conclude that the primary effect responsible for the $r_p$ increase comes from the $\alpha$ core deviating from the center of mass of $^6$He, which is also supported by our calculated proton-neutron and neutron-neutron rms separations. In the future, we are also interested in exploring other spatial correlation operators, such as $r^4$ and $r^6$, to provide a richer physical picture of the structure of light nuclei, constructing semiclassical models capable of presenting detailed structural information of a nucleus, and improving the extrapolation techniques for the long-range operators that are used to compute spatial observables.

\begin{acknowledgments}
We wish to acknowledge useful discussions with Pieter Maris and Mark Caprio. This work was supported in part by the US Department of Energy (DOE) under Grant No.~DE-FG02-87ER40371, No.~DE-SC0018223 (SciDAC-4/NUCLEI), No.~DE-SC0023495 (SciDAC-5/NUCLEI), No.~DE-SC0023707 (Quantum Horizons---NuHaQ), and No.~DE-FG02-95ER40934. This work was performed in part under the auspices of the U.S. Department of Energy by Lawrence Livermore National Laboratory under Contract No.~DE-AC52-07NA27344. Computational resources were provided by the National Energy Research Scientific Computing Center (NERSC), which is supported by the US DOE Office of Science under Contract No. DE-AC02-05CH11231 using NERSC award NP-ERCAP0020944. T.F. thanks the hospitality of the Department of Physics and Astronomy of Iowa State University and Fulbright Brazil for the financial support during his visits in 2018 and 2019. T.F. also thanks the National Council for Scientific and Technological Development (CNPq) Grant No.~306834/2022-7, the S\~ao Paulo Research Foundation (FAPESP) Grant No.~2023/13749-1 and No. 2024/17816-8, and the INCT-FNA Project No.~408419/2024-5. P.Y. is supported by Natural Science Foundation (Henan, China)
under Grant No.~252300421486. We thank the Institute for Nuclear Theory at the University of Washington for its kind hospitality and stimulating research environment. This research was supported in part by the INT's U.S. Department of Energy Grant No.~DE-FG02-00ER41132.
\end{acknowledgments}

\appendix
\section{Rotationally invariant operator with given two-body spin and given pair of isospin $z$ projections}\label{rotscal}

We consider matrix elements of local two-body operators of the form $O(r)P_S$, that is, a rotationally invariant (scalar) operator depending only on the relative spatial coordinate $r$ of the two nucleons, in combination with the two-body spin projection operator $P_S$, where $\langle S_1 \vert P_S \vert S_2 \rangle = \delta_{SS_1}\delta_{SS_2}$.  We adopt the shorthand $\mathcal{O}(r)[S]\equiv O(r) P_S$, and below we focus on operators $\mathcal{O}(r)[S,t_{za},t_{zb}]$ which also project on given isospin $z$ components $t_{za}$ and $t_{zb}$ of the two nucleons.

Since the $O(r)$ operator is a rank-0 tensor, from the Wigner-Eckart theorem,
$
\langle \alpha\beta;J'M'|O(r)|\gamma\delta;J M\rangle
=\delta_{J'J}\delta_{M'M}\langle \alpha\beta;J||O(r)||\gamma\delta;J \rangle,
$
where, in our convention, the double-bar (reduced) matrix element does not factor out $(2J+1)^{-\frac12}$ and remains consistent with the Wigner-Eckart theorem formulated in Eq.~(4.15) of Ref.~\cite{brink1968angular}. 

To select the states of given two-body spin, one can resolve quantities in terms of two-body spin $S$ ($\vec{S}=\vec{s}_\alpha+\vec{s}_\beta$) and two-body total orbital angular momentum $L$ ($\vec{L}=\vec{l}_\alpha+\vec{l}_\beta$) by inserting
\begin{equation}\label{identity}
\mathds{1}=\sum_{L S}|n_\alpha l_\alpha n_\beta l_\beta L, \frac12 \frac12 S;J\rangle \langle n_\alpha l_\alpha n_\beta l_\beta L, \frac12 \frac12 S;J|
\end{equation}
into $\langle \alpha\beta;J||O(r)||\gamma\delta;J \rangle$ twice, where the spin quantum numbers ($\frac{1}{2}$) are explicit while the isospin quantum numbers are omitted for simplification of the notation. %

We find that the reduced matrix element of $\mathcal{O}(r)[S]$ can be expressed as
\begin{widetext}
\begin{equation}\label{lsjj}
\begin{split}
&\langle \alpha\beta;J||\mathcal{O}(r)[S]||\gamma\delta;J\rangle %
\\=&
\smashoperator{\sum_{L_1 L_2 S_1S_2}}
\langle \alpha\beta;J|n_\alpha l_\alpha n_\beta l_\beta L_1, \frac12 \frac12 S_1;J\rangle 
\langle n_\alpha l_\alpha n_\beta l_\beta L_1, \frac12 \frac12 S_1;J||O(r) ||n_\gamma l_\gamma n_\delta l_\delta L_2, \frac12 \frac12 S_2;J\rangle 
\langle n_\gamma l_\gamma n_\delta l_\delta L_2, \frac12 \frac12 S_2;J|\gamma\delta;J \rangle
\\&\times\delta_{SS_1}\delta_{S_1S_2}
\\=&\sum_{L }\langle n_\alpha l_\alpha n_\beta l_\beta L||O(r)||n_\gamma l_\gamma n_\delta l_\delta L\rangle
\langle \alpha\beta;J|n_\alpha l_\alpha n_\beta l_\beta L, \frac12 \frac12 S;J\rangle
\langle n_\gamma l_\gamma n_\delta l_\delta L, \frac12 \frac12 S;J|\gamma\delta;J\rangle,
\end{split}
\end{equation}
where in the last line we have used the fact that $O(r)$ is rotationally invariant, so $L$ is invariant upon the action of $O(r)$, i.e., $\langle L_1 |O(r)|L_2\rangle=\langle L_1|O(r)\delta_{L_1L_2}|L_2\rangle$. We now drop the subscript of $L$ for notational simplicity. There is no need to label $J$ values in the reduced matrix elements, because $O(r)$ only produces a dependence on the radial and orbital quantum numbers.

We identify the ``$LS-jj$'' coupling coefficient \cite{condon1951theory,Edmonds+2016}, i.e., the transformation coefficient between Russell-Saunders coupling and the spin-orbit coupling in Eq.~(\ref{lsjj}), by first expanding the notation 
\begin{equation}
\begin{split}
\langle \alpha\beta;J|n_\alpha l_\alpha n_\beta l_\beta L, \frac12 \frac12 S;J\rangle = &\, \langle n_\alpha l_\alpha \frac12 j_\alpha, n_\beta l_\beta \frac12 j_\beta;J|n_\alpha l_\alpha n_\beta l_\beta L, \frac12 \frac12 S;J\rangle\\
 =&\, \langle l_\alpha \frac12 j_\alpha, l_\beta \frac12 j_\beta;J|l_\alpha l_\beta L, \frac12 \frac12 S;J\rangle\\
 =& \,\sqrt{(2j_\alpha + 1)(2j_\beta + 1)(2L + 1)(2S + 1)}
\begin{Bmatrix}
l_\alpha & l_\beta & L\\
1/2 & 1/2 & S\\
j_\alpha & j_\beta & J
\end{Bmatrix},
\end{split}
\end{equation}
where the curly brackets represent the 9$j$ symbols, which is another way to write an assembly of Clebsch-Gordan coefficients. 

One then obtains
\begin{equation}
\begin{split}
\langle \alpha\beta;J||\mathcal{O}(r)[S]||\gamma\delta;J \rangle =&\langle n_\alpha l_\alpha n_\beta l_\beta L||O(r)||n_\gamma l_\gamma n_\delta l_\delta L\rangle
\\&\times
(2S+1)\sqrt{(2j_\alpha + 1)(2j_\beta + 1)(2j_\gamma + 1)(2j_\delta + 1)}
\\&\times
\sum_{L }(2L+1)
\begin{Bmatrix}
l_\alpha & l_\beta & L\\
1/2 & 1/2 & S\\
j_\alpha & j_\beta & J
\end{Bmatrix}
\begin{Bmatrix}
l_\gamma & l_\delta & L\\
1/2 & 1/2 & S\\
j_\gamma & j_\delta & J
\end{Bmatrix}.
\end{split}
\end{equation}

To project this two-body operator on a fixed pair of isospin $z$ components, we define 
$
\mathcal{O}(r)[S,t_{za},t_{zb}]=O(r)P_SP_{t_{za},t_{zb}}
$, with
\begin{equation}\label{br2}
\begin{split}
\langle \alpha\beta;J||\mathcal{O}(r)[S,t_{za},t_{zb}]||\gamma\delta;J \rangle
=&\langle \alpha\beta;J||\mathcal{O}(r)[S,t_{za},t_{zb}]\delta_{t_{z\alpha},t_{z\gamma}}\delta_{t_{z\beta},t_{z\delta}}||\gamma\delta;J \rangle\\
=&\langle \alpha\beta;J||\mathcal{O}(r)[S]||\gamma\delta;J \rangle \delta_{t_{z\alpha},t_{z\gamma}}\delta_{t_{z\beta},t_{z\delta}}\,g(t_{z\alpha},t_{z\beta})_{t_{za}\leq t_{zb}}.
\end{split}
\end{equation}
 
The resulting matrix element becomes
\begin{equation}\label{brJM}
\begin{split}
\langle \alpha\beta;J'M'|\mathcal{O}(r)[S,t_{za},t_{zb}]|\gamma\delta;J M\rangle =&\,\delta_{t_{z\alpha},t_{z\gamma}}\delta_{t_{z\beta},t_{z\delta}}\delta_{J'J}\delta_{M'M}\,g(t_{z\alpha},t_{z\beta})_{t_{za}\leq t_{zb}}\\
&\times(2S+1)\sqrt{(2j_\alpha + 1)(2j_\beta + 1)(2j_\gamma + 1)(2j_\delta + 1)}\\
&\times\sum_{L }(2L+1)
\begin{Bmatrix}
l_\alpha & l_\beta & L\\
1/2 & 1/2 & S\\
j_\alpha & j_\beta & J
\end{Bmatrix}
\begin{Bmatrix}
l_\gamma & l_\delta & L\\
1/2 & 1/2 & S\\
j_\gamma & j_\delta & J
\end{Bmatrix}\langle n_\alpha l_\alpha n_\beta l_\beta L ||O(r)||n_\gamma l_\gamma n_\delta l_\delta L \rangle.
\end{split}
\end{equation}
\end{widetext}

\section{Harmonic oscillator matrix elements}\label{appendix_ho_matrix}

One can obtain the matrix elements of operators $O(r)=r^0$ and $O(r)=r^2$ in the convenient HO basis with the knowledge that the angular parts of the wavefunctions are not affected, so only matrix elements with $l=l'$ are nonzero:
\begin{equation}
\langle n'l'| r^0 |nl\rangle =\delta_{n,n'}\delta_{l,l'},
\end{equation} 
\begin{equation}
\begin{split}
&\langle n'l'| r^2 |nl\rangle \\=&\,b^2
\left[(2n+l+\frac32)\delta_{n,n'}-\sqrt{n(n+l+\frac12)}\delta_{n,n'+1}\right.\\
&\left.-\sqrt{(n+1)(n+l+\frac32)}\delta_{n,n'-1} \right]\delta_{l,l'},
\end{split}
\end{equation}
where $b\equiv \sqrt{\hbar/(\mu\Omega)}$ with the reduced mass ${\mu}=m_\text{N}/2$ ($m_\text{N}=938.92$~$\text{MeV}/c^2$ is the average mass of a neutron and a proton with their mass given by PDG~\cite{PDG2020}). The derivation of these matrix elements has been detailed in the Appendix of Ref.~\cite{Basili:2019utx}.

\section{Moshinsky transformation}\label{appendix_moshinsky}

We follow the Moshinsky transformation as discussed in Ref.~\cite{PhysRevC.15.423}. The single-particle HO wave function in the two-body relative frame is
\begin{equation}
\langle \Vec{r}|n l m\rangle.
\end{equation}

The two-body oscillator wave function with two-body total orbital angular momentum $L$ and its projection $\lambda$ reads
\begin{equation}
\begin{split}
&\langle \Vec{r}_1 \Vec{r}_2 | n_1 l_1 n_2 l_2 L \lambda\rangle\\=&R_{n_1l_1}(r_1)R_{n_2l_2}(r_2)
\smashoperator{\sum_{m_1m_2}}C
\begin{smallmatrix}
l_1 & \!\!l_2 & \!\!L \\
m_1 & \!\!m_2 & \!\!\lambda
\end{smallmatrix}
Y_{l_1m_1}(\mathbf{\Omega}_1)Y_{l_2m_2}(\mathbf{\Omega}_2).
\end{split}
\end{equation}

Noting that the two-body relative coordinates
\begin{equation}
\Vec{r}=\frac{1}{\sqrt{2}}(\Vec{r}_1-\Vec{r}_2)  \text{,} \quad \Vec{R}=\frac{1}{\sqrt{2}}(\Vec{r}_1+\Vec{r}_2),
\end{equation}
and introducing the two-body center-of-mass orbital angular momentum $\mathcal{L}$ and its projection $\mathcal{M}$ and the center-of-mass radial quantum number $\mathcal{N}$, one arrives at
\begin{equation}
\begin{split}
\langle \Vec{r} \Vec{R} | n_1 l_1 n_2 l_2 L \lambda\rangle= &\sum_{l\mathcal{L}}\psi_i(r,R)\\
&\times\sum_{m\mathcal{M}}C
\begin{smallmatrix}
l & \!\!\mathcal{L} & \!\!L \\
m & \!\!\mathcal{M} & \!\!\lambda
\end{smallmatrix}
\,Y_{lm}(\mathbf{\Omega}_r)Y_{\mathcal{L}\mathcal{M}}(\mathbf{\Omega}_R),
\end{split}
\end{equation}
where $i\equiv \{n_1 l_1 n_2 l_2 l \mathcal{L} L\}$, and
\begin{equation}
\psi_i(r,R) = \sum_{n\mathcal{N}}(nl,\mathcal{N}\mathcal{L};L|n_1 l_1,n_2 l_2 ;L)_1R_{nl}(r)R_{\mathcal{N}\mathcal{L}}(R),
\end{equation}
with the well-known Moshinsky brackets $(nl,\mathcal{N}\mathcal{L};L|n_1 l_1,n_2 l_2 ;L)_1$ subscripted by the ratio of the mass of the two bodies (see e.g., Ref.~\cite{SOTONA197253}).

\vspace{0.7cm}

%


\end{document}